\documentclass[twocolumn,pra,showpacs]{revtex4}
\usepackage{amsmath,amsfonts}
\begin{document}
\draft
\title{\bf Shielding of an external oscillating electric field inside atoms}
\author{V.V. Flambaum
$^{1,2}$
} 
\affiliation{
$^1$
School of Physics, University of New South Wales,  Sydney 2052,  Australia}
\affiliation{$^2$Johannes Gutenberg-Universit\"at Mainz, 55099 Mainz, Germany}
\date{\today}

\begin{abstract}
According to the Schiff theorem an external electric field vanishes at atomic nucleus in a neutral atom, i.e. it is completely shielded
by electrons. This makes a nuclear electric dipole moment (EDM)  unobservable. In this paper an extension of the Schiff theorem to an oscillating electric field is considered. Such field can reach the nucleus and interact with the nuclear EDM .
The enhancement effect appears if the field is in resonance with atomic or molecular transition. The shielding by electrons strongly affects low-energy nuclear electric dipole transition amplitudes in different nuclear reactions including radiative transitions, radiative nucleon capture, photo or electro-excitation of nuclei and  laser-induced or laser-enhanced nuclear reactions.
 \end{abstract}

\maketitle
{\bf Introduction: EDM}. Existence of electric dipole moments (EDM) of elementary particles, nuclei, atoms and molecules in a state with a definite angular momentum violates time reversal invariance (T) and parity (P). EDM also violates CP invariance if the CPT invariance holds. A very extensive experimental and theoretical activity related to EDM is motivated by the need to test unification theories predicting T, P, and CP violation. 

A measurement of the nuclear EDM could  provide information about T,P-odd nuclear forces, neutron and proton EDM.   However, there is a problem here.   A homogeneous static electric field does not accelerate neutral atom. This means that the total electric field ${\bf E}$ acting on the atomic nucleus is zero  since otherwise the charged nucleus would be accelerating, i.e. the external  field is completely shielded by atomic electrons.  The absence of the electric field means that the nuclear EDM $d$ is unobservable, ${\bf d \cdot E}=0$. One may also say that the nuclear EDM is shielded by the atomic electrons and the atomic EDM is zero even if the nucleus has EDM. 

A quantum-mechanical derivation of this result for an arbitrary non-relativistic system of point-like charged particles with EDMs has been done by Schiff  \cite{Schiff}. Schiff also mentioned that his  theorem is violated by the finite nuclear size. The effect of the finite nuclear size was implemented as the nuclear Schiff moment which was introduced in Refs. \cite{Sandars,Hinds,SFK,FKS1986}. An electrostatic interaction between the nuclear Schiff moment and  electrons produces atomic and molecular EDM. Refs. \cite{Sandars,Hinds} calculated  the finite nuclear size effect of the proton EDM.   Refs. \cite{SFK,FKS1986} calculated (and named) the nuclear Schiff moment  produced by the P,T-odd nuclear forces. It was shown in \cite{SFK} that the contribution of the P,T-odd forces to the nuclear EDM and Schiff moment is $\sim 40$ times larger than the contribution of the nucleon EDM.   

The suppression factor for  the atomic EDM relative to the nuclear EDM, proportional to a very small ratio of the squared nuclear radius to the squared atomic radius,  is partly compensated by the factor $Z^2 R_S$, where $Z$ is the nuclear charge and $R_S$ is the relativistic factor \cite{SFK}.   However, even in heavy atoms the atomic EDM is $\sim 10^{3}$ times smaller than the nuclear EDM. An additional 2-3 orders of magnitude enhancement appears in nuclei with the octupole deformation \cite{Auerbach} however such nuclei (e.g. $^{225}$Ra)  are unstable.
 
  The Schiff theorem is also violated by the magnetic interaction \cite{Schiff,Khriplovich}. Corresponding atomic EDMs produced by the nuclear EDM and electron-nucleus magnetic interaction have been calculated in Ref. \cite{FGP}. In light atoms  this mechanism of atomic EDM dominates but in heavy atoms it is smaller than the effect of the finite nuclear size since the latter very rapidly increases with the nuclear charge, as $Z^2 R_S$, while the magnetic effect increases slower, as $Z R_M$ where $R_M$ is the relativistic factor for the magnetic effect \cite{FGP} .
 
There is no complete shielding in ions. For example, in a molecular ion the shielding factor for the nuclear EDM is $(Z_i/Z)(M_n/M_m)$, where $Z_i$ is the ion charge, $Z$ is the nuclear charge, $M_n$ is the nuclear mass and  $M_m$ is the molecular mass \cite{FlambaumKozlov}. Recently the measurement  in the  ionic  molecule HfF$^+$ was performed in Ref.\cite{Cornell}. However, they measured electron EDM which does not have such shielding factor and actually is strongly enhanced in polar molecules \cite{Sandars1965,Flambaum,SushkovFlambaum}. 

There is another interesting feature of the HfF$^+$ experiment \cite{Cornell}. To keep the charged molecule in the trap the authors have to use an oscillating electric field. This is not important for the electron EDM measurement since the electron EDM is not shielded. However, for the nuclear EDM the oscillating field makes the shielding incomplete and the difference with the static case may be  important. Indeed, the interval between the opposite parity rotational levels $\delta E$ in molecules is very small (especially in the case of $\Omega$ doublets formed by the non-zero electron angular momentum projection $\Omega$ on the molecular axis; for   HfF$^+$ $\Omega=\pm 1$ ), and the non-zero frequency effect for $\omega \sim \delta E/\hbar$ should be considered.

{\bf Introduction: nuclear reactions.} Shielding of an external electric field by electrons strongly affects low-energy nuclear electric dipole transition amplitudes. This may happen in  low-energy radiative transitions, radiative nucleon capture, photo or electro-excitation of nuclei and in the laser-induced or laser-enhanced nuclear reactions.

Activity in the latter field has been motivated by the theoretical papers \cite{Zaretski1,Zaretski2} where the laser-induced s-wave neutron capture to a p-wave resonance has been suggested. Capture of a low energy (e.g. thermal) neutron to a p-wave resonance is kinematically suppressed $10^6$ times but the laser field allows an unsuppressed s-wave neutron capture to the p-wave resonance (note that such kinematic enhancement,  combined with  the enhanced mixing of close s- and p-wave compound states (resonances) by the weak interaction, leads to a $10^6$ enhancement of parity violating effects in neutron reactions predicted in Ref. \cite{SF1980}, confirmed in experiment \cite{Alfimenkov} and then studied in numerous experiments involving a hundred of  p-wave resonances in many nuclei - see reviews \cite{SF1982,Gribakin,Bowman}).  

These works initiated an intensive theoretical and experimental activity - see e.g. numerous references in \cite{Ho,Dzyublik}. However, in a striking contrast to the success in the study of the enhanced  parity violating effects in p-wave resonances,  experiments with the laser field  \cite{exp1,exp2,exp3} failed to find the predicted effect. Note that these  theoretical predictions  have not taken into account the electron shielding of the laser field and therefore overestimated the effect.

  The availability of new high power lasers and a significant increase of their frequency range due to an efficient  method of the high harmonic generation (atomic antenna   mechanism \cite{Kuchiev,Corkum})  provide an incentive for a proper account of the electron shielding effect which will be done in the present work.  

{\bf Shielding theory: non-resonant oscillating electric field.}     In our  paper \cite{Dzuba} the shielding  of an external electric field in an ion described by  the relativistic Dirac Hamiltonian for atomic electrons has been considered. It was demonstrated  that the Schiff theorem for the nuclear EDM is still valid both in the "exact" Dirac equation treatment and in the Dirac-Hartree-Fock approximation if the external electric field is included in the self-consistent equations.  This allowed us to perform the Dirac-Hartree-Fock numerical calculations for a static electric field and for an oscillating electric field in Tl$^+$. 

The screened field $E=E_0 +<E_e>$ oscillates in space, has a maximal magnitude $E \approx -3 E_0$ near the radius of the 1$s$ shell, $r=a_B/Z$ ($a_B$ is the Bohr radius),  and becomes very small near the nucleus. It was concluded that the deviation of the electric field at the nucleus from zero in a neutral system is proportional to $\omega^2$, where $\omega$ is the electric field oscillation frequency.   However, there was no formula derived for the shielding factor in the case of the oscillating field. The aim of the present paper is to derive such formula and extend the Schiff theorem to the case of the oscillating electric field.

The Hamiltonian of an atom in an external electric field along the $z$-axis $E_z=E_0 \cos(\omega t)$ may be presented as 
\begin{eqnarray}\label{HE}
H_E=H_0 - E_z D_z \,, \\
D_z=e \sum_{k=1}^{N} z_k\,,
\end{eqnarray}
where $H_0$ is the Schrodinger or the Dirac Hamiltonian for the atomic electrons in the absence of the external field $E_z$, $N$ is the number of the electrons, $Z_i=Z-N$,  $e=-|e|$ is the electron charge, $z_k$ is the $z$-axis projection of the electron position relative to the nucleus. We assume that the nuclear mass is infinite and neglect very small effects of the    Breit and  magnetic interactions. The electric field on the nucleus  may be presented as $E_n=(E_0 +<E_e>) \cos(\omega t)$, where the electron electric field on the nucleus is
\begin{equation}\label{Ee}
E_e= -
e\sum_{k=1}^{N}\frac{z_k}{r_k^3}=-\frac{i}{Z e\hbar}[P_z,H_0] \,,
\end{equation}
where $P_z=\sum_{k=1}^{N}p_{z,k}$ is the total momentum of the atomic electrons. The second equality follows from the differentiation of the nuclear Coulomb potential in the Dirac or Shrodinger Hamiltonian $H_0$ since the total electron momentum $P_z$ commutes with the electron kinetic energy and the electron-electron interaction. Using the time dependent perturbation theory \cite{Landau} for the oscillating  perturbation $D_zE_z$ we obtain
\begin{eqnarray}\label{HE}
\nonumber
<E_e>= - E_0 \sum_n \frac{(\epsilon_0 - \epsilon_n)}{(\epsilon_0 - \epsilon_n)^2 -\epsilon^2}\times\\
\nonumber
 (<0|E_e|n><n|D_z|0> +<0|D_z|n><n|E_e|0>)=\\
 \nonumber
 - \frac{iE_0}{Z e\hbar}\sum_n \frac{(\epsilon_0 - \epsilon_n)^2}{(\epsilon_0 - \epsilon_n)^2 -\epsilon^2}\times\\
 \nonumber
 (<0|P_z|n><n|D_z|0> - <0|D_z|n><n|P_z|0>)\,.\\
\end{eqnarray}
The second equality follows from Eq. (\ref{Ee}) and the relation $<0|[P_z,H_0]|n>= (\epsilon_0 - \epsilon_n)<0|P_z|n>$, $\epsilon =\hbar \omega$. The energy dependent factor  may be presented as
\begin{equation}\label{energy}
\frac{(\epsilon_0 - \epsilon_n)^2}{(\epsilon_0 - \epsilon_n)^2 -\epsilon^2}=1 + \frac{\epsilon^2}{(\epsilon_0 - \epsilon_n)^2 -\epsilon^2}\,.
\end{equation}
The energy independent  term 1 in the right hand side allows us to  sum over states $|n>$ in Eq. (\ref{HE}) using the closure and then use the commutator relation $[P_z,D_z]= -i e \hbar N$.  The result is
\begin{eqnarray}\label{shielding}
\nonumber
<E_e>= -E_0 \frac{N}{Z} - 
 \frac{iE_0}{Z e\hbar}\sum_n \frac{\epsilon^2}{(\epsilon_0 - \epsilon_n)^2 -\epsilon^2}\times\\
 \nonumber
 (<0|P_z|n><n|D_z|0> - <0|D_z|n><n|P_z|0>)\,.\\
\end{eqnarray}
  Using the non-relativistic commutator relation  $P_z=\frac{i m} {e \hbar} [H_0,D_z]$ (here $m$ is the electron mass)  we can express the induced electron field on the nucleus in terms of the atomic dynamical polarisability $\alpha_{zz}(\omega)$: 
\begin{eqnarray}\label{polarizability}
\nonumber
<E_e>= -E_0 \frac{N}{Z} -
 E_0 \alpha_{zz}\frac{\epsilon^2 m}{Z e^2\hbar^2}\,,\\
 \alpha_{zz}=2 \sum_n \frac{(\epsilon_n - \epsilon_0)<0|D_z|n>^2}{(\epsilon_n - \epsilon_0)^2 -\epsilon^2}\,.
\end{eqnarray}
The values of the dynamical polarizabilities are measured and calculated for many atoms, they appear in the expression for the refractive index. There are high precision computer codes for the calculations of the dynamical polarizabilities, see e.g. \cite{DzubaLa,Schiller}.

 It may be instructive to present the  formula for the total electric field amplitude $E_t$ at the nucleus using the energy and the polarizabilty in atomic units,  ${\tilde \epsilon} = \frac{\epsilon}{e^2/a_b}$ and ${\tilde  \alpha}_{zz}=\frac{ \alpha_{zz}}{a_b^3}$:   
 \begin{equation}\label{polarizabilityAtomic}
E_t= 
E_0 ( \frac{Z_i}{Z} - \frac{    {\tilde \epsilon}^2 {\bf \tilde  \alpha}_{zz}}{Z})
\end{equation}
If $\epsilon^2=(\hbar \omega)^2  \ll  (\epsilon_0 - \epsilon_n)^2$ we have the static-type screening of the external field,  $E_0+E_e=E_0 (1-N/Z)=E_0 Z_i/Z$, i.e. the complete shielding of the external field in neutral systems where the ion charge $Z_i=Z-N=0$.  

The shielded field  is proportional to $1/Z$, so it may seem that the shielding is stronger in heavy atoms. However, it is not necessary the case since in hydrogen and helium  ${\bf \tilde  \alpha}_{zz} \sim 1$ while in caesium ($Z$=55) ${\bf \tilde  \alpha}_{zz} \sim 400$.  Indeed, the numerical value  of the polarizability ${\tilde  \alpha}_{zz}$ in atomic units often exceeds the value of the nuclear charge $Z$, therefore, the suppression of the field mainly comes from the frequency of the field oscillations in atomic units,  ${\tilde \epsilon}$. 

As an illustration, let us consider a numerical example. One of the largest parity violating effects (7 \%) have been observed in the 0.734 eV p-wave resonance in $^{139}$La, $Z=70$.  This means that the kinematic factor and the  mixing of s and p compound states by the weak interaction are large.  Therefore, it looks natural to use this resonance to search for the capture of neutron in a laser field which also may provide mixing of the s and p compound states and enhance capture of neutron to the p-wave resonance.

The static scalar polarizabilty of La is  ${\bf \tilde  \alpha}_{s} = 213.7$ \cite{DzubaLa}. Thus, in a low frequency laser field, $\tilde \epsilon=1/27.2$ (1 eV), the shielding factor  is   0.005. However, it rapidly increases with $\tilde \epsilon$ and reaches the pole of ${\bf \tilde  \alpha}_{zz}$ at the position of the La atom energy level  $\tilde \epsilon=0.0604$ (1.64 eV).

{\bf Atomic resonance.} When the frequency increases and approaches the resonance, $\epsilon^2=(\hbar \omega)^2  \approx (\epsilon_0 - \epsilon_n)^2$, the induced electron field may become much larger than the external field amplitude $E_0$. The field remains finite for $\epsilon^2=(\epsilon_0 - \epsilon_n)^2$ due to the widths of the excited states  which should be added to the energy denominators (where we should have $ \epsilon_n -i \Gamma_n/2$ instead of $ \epsilon_n$). 

 If the width is small ($\Gamma_n \ll e E_0 <0|D_z|n>$) and may be neglected, the Rabi oscillations between the two resonating states happen (electron oscillates between the ground state and excited state and  at any instant the wave function is a superposition of two states). 
 
 The solution for a two-level system with energies $E_0$ and $E_n$ subjected to a periodic perturbation is presented  in the textbook  \cite{Landau}. We just should calculate the electron field $E_e$ using this two-state wave function. Use of the commutator relations in  Eq. (\ref{Ee}), $<0|[P_z,H_0]|n>= (\epsilon_0 - \epsilon_n)<0|P_z|n>$ and $P_z=\frac{i m} {e \hbar} [H_0,D_z]$ 
 leads to the following expression for the resonance contribution to the electric filed at the nucleus for $\epsilon^2=(\hbar \omega)^2 = (\epsilon_0 - \epsilon_n)^2$:
\begin{eqnarray}\label{resonance}
<E_e>= E_r \sin(\Omega t)\sin(\omega t)\,, \\
\Omega=2 e E_0 <0|D_z|n>/\hbar\,,\\
E_r=\frac{{\tilde \epsilon}^2{\tilde D}_z}{Z} \frac{e}{a_B^2}= \frac{{\tilde \epsilon}^2{\tilde D_z}}{Z} \,\, \times 5.14 \times10^9 {\rm V/cm} \,,
\end{eqnarray}
where ${\tilde D}_z=\frac{<0|D_z|n>}{e a_B}$. The frequency  of these Rabi oscillations $\Omega$ is determined by the strength of the external field  $E_0$  but the field on the nucleus  does not depend on $E_0$ and is defined by the electron field which has a scale $e/a_B^2 =5.14 \times 10^9$ V/cm.  

Again, the suppression of the field at the nucleus  $ \sim {\tilde \epsilon}^2$ appears if the field oscillation frequency is small.  This may be  the case if we want to use a resonance between close  opposite parity levels (e.g. in a molecule) to measure the nuclear EDM. The frequency of the oscillations should not be too high since one has to separate the oscillating signal at this frequency. For example, one may rotate the nuclear spin  in a unison with the rotating electric field. Using a very optimistic estimate    $ {\tilde \epsilon}=10^{-4}$ (658 GHz)  and $Z \sim 1$ we obtain $E_r \sim  50$ V/cm.  This field does not look  large, and the experiment itself looks too complicated to do.  

 However, there are two arguments in favour of such attempt. Nuclear EDM in light nuclei such as $^1$H, $^2$H and $^3$He may be calculated more reliably than the Schiff moment in heavy nuclei. Indeed, the formula for the Schiff moment contains two terms of opposite signs (the second term comes from the electron shielding  effect).  As a result, the sophisticated  many-body calculations for $^{199}$Hg \cite{Engel} failed to predict the magnitude and even the sign of the Schiff moment.    
 
 The second argument is that the static effective field (the screened external field which is not zero due to the magnetic interaction) for  the nuclear EDM in light atoms is very small.
According to \cite{Schiff} the suppression factor for $^3$He is 10$^{-7}$, i.e. the external filed 30 KV/cm corresponds to the effective field acting on the $^3$He EDM of only  0.003 V/cm.
Therefore, if someone would decide to do a  measurement of a theoretically "clean" light nucleus EDM in a neutral molecule or atom, an oscillating electric field is possibly not the worst option.  

As an example of the strong field at the nucleus,  we take the lanthanum resonance case  $Z=57$,  $\tilde \epsilon=0.0604$ (1.64 eV) and  use a rough estimate ${\tilde D}_z \sim 0.3$. This gives the field at the nucleus $E_r \sim 10^5$ V/cm. In the higher-energy resonances the field may be an order of magnitude larger due to the larger $\tilde \epsilon ^2$ and ${\tilde D}_z$. Such very strong field gives an incentive to study laser-induced nuclear reactions using atomic resonances, for example, to repeat the laser-induced neutron capture experiments \cite{exp1,exp2,exp3}.

 
    This work is supported by the Australian Research Council and Gutenberg Felowship. The author is grateful to Wick Haxton  for a discussion of the Schiff theorem effect on the nuclear transitions, to Vladimir Gudkov for sending me Ref. \cite{Ho}, to Vladimir Dzuba for the numerical tests of the formula  (\ref{polarizabilityAtomic}) and to Igor Samsonov for valuable discussions.

\end{document}